\documentclass[twocolumn,english,prl,notitlepage,nofootinbib,floatfix,superscriptaddress]{revtex4-2}
\usepackage[T1]{fontenc}
\usepackage[utf8]{inputenc}
\usepackage{refstyle}
\usepackage{amsmath}
\usepackage{bm} % for bold symbols in math mode
\usepackage{amssymb}
\usepackage{graphicx}
\usepackage[pdftex, hidelinks]{hyperref}
\usepackage[usenames,dvipsnames]{xcolor}
\usepackage{bbm}
\usepackage{tikz}
\usepackage{xspace}
\usepackage{bm}
\usepackage{booktabs}
\usepackage{babel}
\usepackage[separate-uncertainty=true]{siunitx}
\usepackage[makeroom]{cancel}
\usepackage{siunitx}

\usepackage[normalem]{ulem} % for crossing out text (ie text strikethrough)
\usepackage{comment}  %% ADDED 13.07.2020
\usepackage{braket}  %% ADDED 14.07.2020
\usepackage{float} %% ADDED 15.07.2020 
\usetikzlibrary{matrix}

\newcommand\myshade{85}
\colorlet{mylinkcolor}{violet}
\colorlet{mycitecolor}{YellowOrange}
\colorlet{myurlcolor}{Aquamarine}
\hypersetup{
  linkcolor  = mylinkcolor!\myshade!black,
  citecolor  = mycitecolor!\myshade!black,
  urlcolor   = myurlcolor!\myshade!black,
  colorlinks = true,
}

\usepackage{soul}
\usepackage{xr}

\DeclareMathAlphabet\mathbfcal{OMS}{cmsy}{b}{n}

\DeclareMathOperator{\sinc}{sinc}

\begin{document}

\title{
Entanglement-enhanced magnetic induction tomography
}

\author{Wenqiang Zheng}
\thanks{These authors contributed equally to this work.}
\affiliation{Niels Bohr Institute, University of Copenhagen, Blegdamsvej 17, 2100 Copenhagen Ø, Denmark}
\affiliation{Present address: Zhejiang Provincial Key Laboratory and Collaborative
Innovation Center for Quantum Precision Measurement, College of Science,
Zhejiang University of Technology, Hangzhou 310023, China}

% $^{1,2}$, Hengyan Wang$^{1,3}$, Rebecca Schmieg$^{1}$, Alan Oesterle$^{1}$ and Eugene S. Polzik$^{1}$}

\author{Hengyan Wang}
\thanks{These authors contributed equally to this work.}
\affiliation{Niels Bohr Institute, University of Copenhagen, Blegdamsvej 17, 2100 Copenhagen Ø, Denmark}
\affiliation{Present address: Department of Physics, Zhejiang University of Science and Technology, Hangzhou 310023, China}

\author{Rebecca Schmieg}
\thanks{These authors contributed equally to this work.}
\affiliation{Niels Bohr Institute, University of Copenhagen, Blegdamsvej 17, 2100 Copenhagen Ø, Denmark}

\author{Alan Oesterle}
\affiliation{Niels Bohr Institute, University of Copenhagen, Blegdamsvej 17, 2100 Copenhagen Ø, Denmark}

\author{Eugene S. Polzik} \email{polzik@nbi.ku.dk}
\affiliation{Niels Bohr Institute, University of Copenhagen, Blegdamsvej 17, 2100 Copenhagen Ø, Denmark}

\begin{abstract}
 Magnetic induction tomography (MIT) is a sensing protocol, exploring conductive objects via their response to radio-frequency magnetic fields. MIT is used in nondestructive testing ranging from geophysics to medical applications. Atomic magnetometers, employed as MIT sensors, allow for significant improvement of the MIT sensitivity and for exploring its quantum limits.
 Here we report entanglement-enhanced MIT with an atomic magnetometer used as the sensing element. We generate an entangled and spin squeezed state of atoms of the sensor by stroboscopic quantum non-demolition measurement. We then utilize this spin state to demonstrate the improvement of one-dimensional MIT sensitivity beyond the standard quantum limit. 

\end{abstract}

\maketitle

% Words: 2043, captions: 397, 7 math line -> 112 words. Not sure if math inside text counted correctly.
% Figures (see at figures) is 230 + 266 + 135 + 110 = 741
%Total 3300 words, however, I am not sure if this includes math inline, if it does, it is another 172 words!!!
% Edit 30.12.22: now 3622 words

Magnetic Induction Tomography (MIT) \cite{griffiths2001magnetic} uses a radio-frequency (RF) magnetic field from a coil to induce eddy currents in an object of interest. Detection of eddy currents allows to reveal information about the composition and shape of an object non-destructively and non-invasively %revealing information about the composition and
%the shape of the object through their 
since the eddy currents depend on the conductivity and permeability.
While the bulk of MIT applications detect the eddy currents using a pick-up coil, atomic RF magnetometers (AM) have been introduced as viable high-sensitivity alternative sensors for MIT \cite{wickenbrock2014magnetic,Deans2016EMinductionimaging,Wickenbrock2016Eddycurrentimaging, Deans2017imgaingAM,Jensen2019a}.
% The eddy currents depend on an object's conductivity and permeability, \blue{their detection reveals} information about the composition and shape of the object non-destructively and non-invasively. %The MIT is a nondestructive and non-invasive method used to study hidden and concealed objects with applications in various kinds of sensing. 

Quantum sensing and metrology is one of the major fields within quantum information technologies \cite{Degen2017QuantumSensing}. It exploits quantum states of light and matter, such as entanglement and squeezing, to improve the sensitivity of sensors beyond the standard quantum limits (SQL), the boundaries existing in the absence of quantum correlations.
Within atomic physics, quantum enhancement of sensitivity has been demonstrated for electric field sensing \cite{gilmore2021quantum}, clock \cite{Derevianko2011, Schioppo2017}, magnetometer \cite{bao2020spin,wasilewski2010quantum,sewell2012magnetic}, and interferometry \cite{Appel2009}.

Here , we demonstrate a novel quantum metrology protocol, quantum-enhanced magnetic induction tomography (QMIT). The protocol exploits {(1)} implementing an anti-Helmholtz coil geometry providing efficient cancellation of classical noise %\sout{of the apparatus {?} }
\cite{Jensen2019a}; {(2)} introducing a stroboscopic measurement sequence at an RF rate suppressing quantum back-action (QBA)%\sout{of the measurement}
; {(3)} generation of spin squeezed states of the atomic spin sensor; {(4)} measuring the eddy current signal compatible with the stroboscopic measurement sequence.

For a sensor containing  $N_\text{A}$ uncorrelated particles, such as atomic spins, the SQL of measurement sensitivity scales as ${1}/{\sqrt{N_\text{A}}}$ \cite{Budker2007}. It is set by the Heisenberg uncertainty principle restricting how precisely two non-commuting operators %for spin projections 
can be measured simultaneously. %Quantum e
%Entanglement and squeezing of the spin state 
Spin squeezing and entanglement can improve the sensitivity beyond the ${1}/{\sqrt{N_\text{A}}}$ limit. While large spin ensembles with $\sqrt{N_\text{A}}\gg1$ promise the most sensitive measurements, technical imperfections growing with $N_\text{A}$ often preclude overcoming %\sout{or even reaching}
the SQL. Reduced technical fluctuations %\sout{at higher frequencies} 
make the SQL more attainable for measurements in the RF range \cite{savukov2007detection}. As  MIT utilizes RF field sensing, quantum enhancement appears to be an attractive
approach for enhancing the sensitivity.%\blue{The last sentence can be removed i.m.o. as it is covered 2 sentences before}

\begin{figure*}[!]
\includegraphics[width=16.5cm]{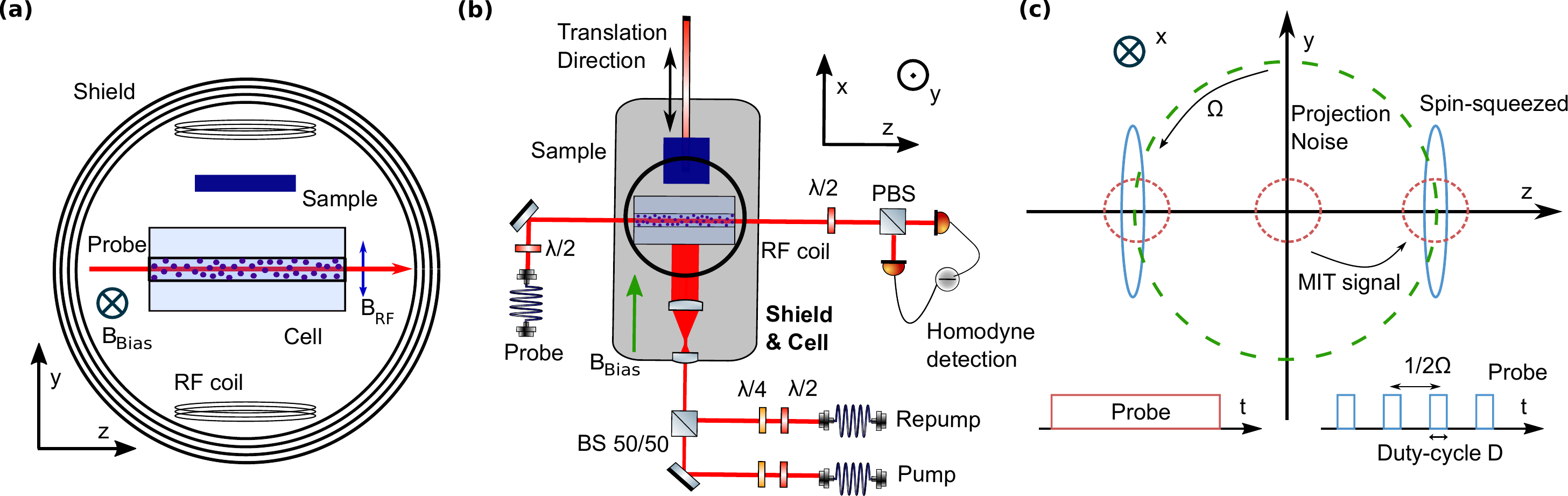}
\caption{Setup for entanglement-enhanced MIT. (a) Configuration of RF coils, the probing direction, the bias magnetic field, the conductive sample and the vapor cell inside the magnetic shield. 
(b) Simplified experimental setup (top view), where $\lambda/2$ and $\lambda/4$  {indicate} half- and quarter-waveplates, (P)BS  {indicates} a (polarizing) beamsplitter. 
(c) Illustration of the trajectory of the spin projection $\protect\overrightarrow{J}_{\bot}$ (dashed green line) in the presence of the MIT signal, together with the  {PN} of the CSS state (red) and the squeezed state (blue). The insets show the time  {sequences} for the continuous and stroboscopic probing.}  
\label{fig: exp. setup}
\end{figure*} % Aspect ratio: 3.2, extrawide, 230 words equivalent

The experimental setup of QMIT is shown in Fig. \ref{fig: exp. setup}. The atomic sensor, containing  $N_\text{A}$ cesium atoms inside a vapor cell, is placed inside a bias magnetic field $B_\text{bias}$ along the $x$-axis corresponding to $\Omega_\text L \approx 725$ kHz  Larmor frequency. The protocol is facilitated by the long transverse spin relaxation time $T_2\approx2.35$ ms which is due to the anti-relaxation coated cell walls \cite{balabas2010high}.
Optical pumping prepares the ensemble in a coherent spin state (CSS) with $m_F=F=4$, for which
$J_x=\left< \hat{J}_x \right> =\sum_{k=1}^{N_\text{A}}{\left< \hat{j}_{x}^{k} \right>}=FN_\text{A}$, where $\hat{j}^k_x$ refers to the $k$-th atom's spin. %of the $k$-th atom. 
The AM is placed in between two anti-Helmholtz RF coils
and monitors the magnetic field response $\overrightarrow{B}_\text{ec}$ generated by the eddy currents induced in the conductive object (Fig. \ref{fig: exp. setup} (a)). 
In the absence of an object, the total RF field is zero at the location of the sensor and hence no transverse spin component is  driven. 
The minimal quantum fluctuation for uncorrelated spins, corresponding to the projection noise (PN)  $ \text{Var}\left( \hat{J}_y \right) =\text{Var}\left( \hat{J}_z \right) =\frac{F}{2}N_\text{A}$ arising from the Heisenberg uncertainty principle, limits the sensitivity of the AM.

In the presence of a conductive object, a non-zero transverse spin component $\overrightarrow{J}_{\bot}$ is created (Fig. \ref{fig: exp. setup} (c)):
\begin{equation}
\left< \overrightarrow{J}_{\bot} \right> = \frac{\gamma }{2}B_\text{ec}J_xT_2\left[ 1-\exp \left( -\tau/{T_2} \right) \right],
\end{equation}
where $\tau$ is the duration of the RF pulse, $T_2$ is the transverse spin coherence time and $\gamma$ is the gyromagnetic ratio.  
Monitoring $\overrightarrow{J}_{\bot}$ 
by polarization homodyne detection (Fig. \ref{fig: exp. setup} (b)) allows extracting information about the induced eddy currents, and hence about the properties of the sample.
With the probe light far detuned from any atomic transition, %(Fig. \ref{fig: exp. setup} (d))
we can realize a quantum nondemolition (QND) measurement of the spin component $\overrightarrow{J}_{z}$ via Faraday interaction  $\hat{H}_\text{F} \propto\frac{\kappa}{\sqrt{N_\text{A}N_\text{P}}}\hat{S}_z\hat{J}_z$ \cite{hammerer2010quantum}, where  $\kappa \propto \sqrt{N_\text{A}N_\text{P}}$ is the coupling constant and $N_\text{P}(N_\text{A})$ is the photon(atom) number. $\hat{S}_z$ is the Stokes operator of the probe light whose value is equal to the difference between right- and left-hand circular polarized components. The Stokes operators obey $[\hat{S}_{z},\hat{S}_{y}]=iS_{x}$, where $S_{x}$ can be treated as a number for input light polarized along $x$-axis.

The homodyne detection 
 {yields} the Stokes operator $\hat{S}_{y}^\text{out}\propto\hat{J}_z=\hat{J}_{z_0}\cos(\Omega t)+ \hat{J}_{y_0}\sin(\Omega t)$. Here $\hat{J}_{z_0}$ and $\hat{J}_{y_0}$ are the spin projections in the rotating frame satisfying $[\hat{J}_{z_0},\hat{J}_{y_0}]=i{J}_{x0}=iFN_A$. 
For continuous probing, $\hat{J}_{z_0}$ and $\hat{J}_{y_0}$ are measured with alternating strength  proportional to $\sin(\Omega t)$ and $\cos(\Omega t)$ per Larmor precession, respectively.  %alternately \red{measured when $\sin(\Omega t)\rightarrow\pm{1}$ and $\cos(\Omega t)\rightarrow\pm{1}$ in every Larmor precession period.} \blue{this sentence indicates that we measure every quarter precession. I think what is confusing to readers is that it is not alternately between them, but the strength of their contribution is alternating with sine and cosine.}
 {Simultaneously}, extraneous quantum back-action noise (BAN) is imprinted onto the conjugate components $ \hat{J}_{z_0}$ and $\hat{J}_{y_0}$ via light-atom interaction \cite{hammerer2010quantum}. In the laboratory frame,  as $\overrightarrow{J}_{\bot}$ rotates around $\overrightarrow{B}_\text{bias}$ at the frequency $\Omega$, BAN is imprinted onto both spin components $\overrightarrow{J}_{y}$ and $\overrightarrow{J}_{z}$, and thus affect the readout noise of the polarization homodyning.
Quantum fluctuations of light,
corresponding to the photon shot noise (SN), also increase the measurement uncertainty.
Therefore, a continuous measurement of a precessing spin suffers from SN, PN and BAN \cite{hammerer2005teleportation}. The total quantum noise of the cosine quadrature of the Stokes component $\left( \hat{S}_{y}^\text{out} \right)$, recorded by a lock-in amplifier (LIA), can be expressed as 
\begin{equation}
\text{Var}\left( \hat{S}_{y,c}^\text{out} \right) \approx \frac{N_\text{P}}{4}\left( 1+\frac{\kappa ^2}{2}+\frac{\kappa ^4}{12} \right),
\end{equation}
where the three terms correspond to SN, PN and BAN, respectively. 
As the signal grows linearly with $\kappa$, the SQL of a continuous measurement is achieved for $\kappa^4 = 12$, which optimizes the signal-to-noise ratio $
\text{SNR}\propto {\kappa/\sqrt{1+\kappa ^2/2+\kappa ^4/12}}$.
The respective SQL of the total noise variance is thus $2(1+\kappa^2/2+\kappa^4/12)/\kappa^2=1+2/\sqrt{3}$ times greater than the projection noise variance and the standard deviation is approximately $1.47$ time greater.

\begin{figure}[!]
\includegraphics[width=8cm]{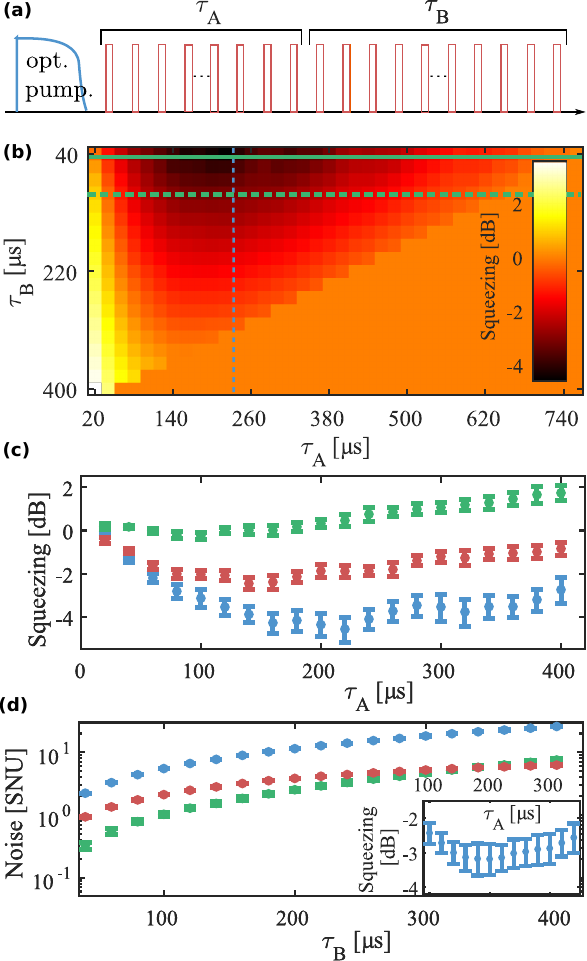}
\caption{Generation of a spin squeezed state of the magnetometer. 
(a) Pulse sequence used for squeezing demonstration consisting of optical pumping and a train of stroboscopic probing pulses modulated at twice the Larmor frequency. %\sout{\red{Squeezing preparation and verification have the same pulse pattern, while their durations $\tau_\text{A}$ and $\tau_\text{B}$ can be varied.} ->info is contained in previous sentence and in text}
(b) Experimental demonstration of squeezing versus preparation duration $\tau_\text{A}$ and verification duration $\tau_\text{B}$ for 15\,\% duty-cycle. Green horizontal line (solid): slice for $\tau_\text{B} = \SI{40}{\micro\second}$ used in subfigure c. Blue vertical line: slice for $\tau_\text{A} = \SI{220}{\micro\second}$ used in subfigure d. Green horizontal line (dashed): slice used for inset in subfigure d.
(c) Achievable squeezing versus $\tau_\text{A}$ for $\tau_\text{B} = \SI{40}{\micro\second}$ for 15\,\% (blue), 50\,\% (red) and 90\,\% (green) duty-cycle.  
(d) Projection noise (red), unconditional variance (blue) and conditional variance (green) versus $\tau_\text{B}$ for optimal $\tau_\text{A} = \SI{220}{\micro\second}$. The noise is normalized to light shot noise units (SNU). The inset shows squeezing for $\tau_\text{B} = \SI{100}{\micro\second}$ versus $\tau_\text{A}$.
The error bars in subfigures (c) and (d) are obtained from statistical analysis of eight data sets, each containing 4\,000 repetitions.
}
\label{fig: squeezdem}
\end{figure} % aspect ratio: 0.61, word equivalent: 266 words

A time-dependent  {measurement}, e.g., a stroboscopic measurement, enables back-action-free measurement of one spin component with a sensitivity exceeding the SQL \cite{ Braginsky1980,Vasilakis2015b,Meng2020}.
For stroboscopic probing at twice the Larmor frequency \cite{Vasilakis2015b}, illustrated in Fig. \ref{fig: exp. setup} (c),  {quantum noise} of the probe observable takes the following form 
\begin{equation}
\text{Var}\left( \hat{S}_{y,c}^\text{out} \right) \approx \frac{\eta N_\text{P}}{4}\left( 1+\frac{\tilde{\kappa}^2}{2}+C\frac{\tilde{\kappa}^4}{12} \right),
\end{equation}
where $\eta =1+\sinc\left( \pi D \right)$, $D$ is the duty-cycle of stroboscopic probing, $\tilde{\kappa}=\sqrt{\eta}\kappa$, and $
C=\frac{1-\sinc\left( \pi D \right)}{1+\sinc\left( \pi D \right)}$.
For a $\delta$-pulse ($D=0$), perfect quantum back-action evasion is achieved, allowing for a QND measurement to be realized. The magnetic sensitivity for the eddy current detection then approaches $\delta B_\text{ec}\propto {1}/{\text{SNR}}\propto {\sqrt{\left( 1+{\tilde{\kappa}^2}/{2} \right)}}/{\tilde{\kappa}}
$.

As a first step, we verify a spin-squeezed state of the atoms contained in an interaction volume of $\SI{500}{\micro\meter}\times\SI{500}{\micro\meter}\times\SI{25}{\milli\meter}$ using the sequence shown in Fig. \ref{fig: squeezdem} (a). 
%Atoms evaporating from the cesium reservoir in the stem fill the channel of size %\sout{within the chip through a micro-hole (?)}
%The probe light propagates along this channel of size . 
%The cell is coated with spin anti-relaxation coating.  %The temperature of the stem defines the vapour pressure and hence the number of atoms interacting with the probe light. 
 {The ensemble consists of $1.5\times 10^{9}$ atoms at the temperature of $ 55~^\circ$C.} Using optical pumping  \cite{SM}, the atoms are prepared in  {the CSS}. %move to supplements: using optical pumping lasers "Pump" and "Repump", locked to the $F=4\rightarrow F' = 4$ on D1-line and $F = 3\rightarrow F'=$ 2,3 crossover transition on the D2-line, respectively. 
Typically, we achieve an atomic polarization of 97.5\,\%, verified by pulsed magneto-optical resonance spectroscopy \cite{SM}. The imperfection leads to the spin projection noise %before the QND measurement 
19.5\,\% higher than that of the CSS \cite{SM}. The spin noise is calibrated using the measured spin noise of the  {unpolarized} atomic ensemble,  {namely} the thermal spin state (TSS), since the TSS is insensitive to classical noise and BAN  \cite{SM}.
 %The light pulses are generated using acousto-optical modulators. 

%Optical pumping, preparing a spin polarization of 97.5 \%, 
Optical pumping is followed by two sequences of stroboscopic probing pulses, modulated at $2\Omega_\text L$ with varying duty cycles, generated using acousto-optical modulators. The probe laser is locked with a detuning of \SI{1.95}{\giga\hertz} from the $F=4\rightarrow F'=$ 4,5 crossover transition of the D2-line.
The first stroboscopic sequence with duration $\tau_\text{A}$ prepares a  squeezed state via QND measurement, while the second sequence with duration $\tau_\text{B}$ verifies the degree of spin squeezing when conditioning on the outcome of the first stroboscopic  measurement.
%of the atomic spin during $\tau_\text{A}$}.
The sequence is repeated thousands of times, allowing  us to estimate $\left< \overrightarrow{J}_{\bot} \right>$.  The signal of each individual sequence is demodulated using an LIA  and then recorded. Here, the outcomes are denoted $Q_\text{A}$ and $Q_\text{B}$ for the squeezing generation and verification processes, respectively. 
Conditioning the signal $Q_\text{B}$ during $\tau_\text{B}$ on the preceding signal $Q_\text{A}$ during $\tau_\text{A}$, allows to determine the conditional variance 
%as in \cite{Vasilakis2015b}.
\begin{equation}
\begin{split}
\text{Var}\left( Q_\text{B}|Q_\text{A} \right) &=\min \left( \text{Var}\left( Q_\text{B}-\alpha Q_\text{A} \right) \right) \\
&=\text{Var}\left( Q_\text{B}-\alpha _\text{opm}Q_\text{A} \right) \\
&=\text{Var}\left( Q_\text{B} \right) -\frac{\text{Cov}^2\left( Q_\text{B},Q_\text{A} \right)}{\text{Var}\left( Q_\text{A} \right)},
\end{split}
\label{eq: cond.var.def}
\end{equation}
where $\alpha$ is the feedback parameter whose optimal value $\alpha _{\text{opm}}=\frac{\text{Cov}\left( Q_\text{B},Q_\text{A} \right)}{\text{Var}\left( Q_\text{A} \right)}$ minimizes the conditional variance.

From the conditional and unconditional variances during $\tau_\text{B}$, we find the degree of  {spin squeezing} as
\begin{align}
\xi ^2 = \frac{\text{Var}\left( Q_\text{B}|Q_\text{A} \right) -\text{SN}_\text{B}-\text{EN}_\text{B}}{\text{Var}\left( Q_\text{B} \right) -\text{SN}_\text{B}-\text{EN}_\text{B}},
\label{eq:squeezing}
\end{align}
where $\text{SN}_\text{B}$ and $\text{EN}_\text{B}$ are photon shot noise and electronic noise contributions during the verifying process.
With the reduced conditional variance, the quantum noise limited  sensitivity increases to  $\delta B_{\text{ec}}\propto {\sqrt{\left( 1+{\xi ^2\tilde{\kappa}^2}/{2} \right)}}/{\tilde{\kappa}}$.

\begin{figure}[!]
\includegraphics[width=8cm]{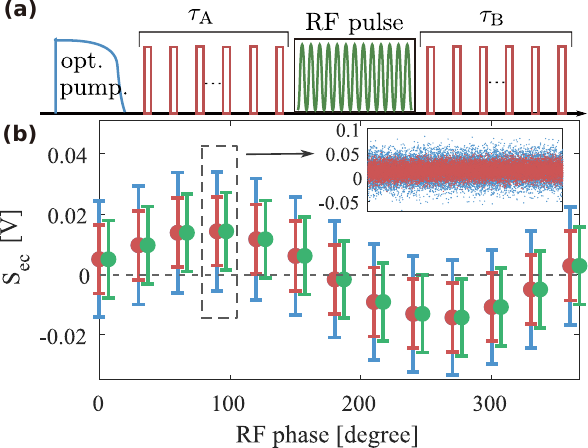}
\caption{Entanglement-enhanced eddy current measurement.  
(a) Pulse sequence with RF pulse between stroboscopic probing pulses. (b) Eddy current signals ($\text{S}_\text{{ec}}$), the difference  {between} sample and background  {signals}, as a function of the RF phase. %obtained by subtracting the background signal from the sample signal. 
Each point is averaged over 16\,000 measurements, while error bars represent single-shot uncertainty. {Blue and red points represent} unconditional and conditional results{, respectively}. The green error bars represent the quantum noise $\sqrt{\text{SN}_\text{B}+ \text{PN}_\text{B}}$, corresponding to the back-action evaded measurement without squeezing, horizontally shifted  for clarity. The inset shows the unconditional (blue) and conditional (red) signal distribution for the RF phase of $90^\circ$. %\blue{should we address in the text that 0 and 360 degree points don't agree completely (drifts, etc.)?}
}
\label{fig: signalVSphase}
\end{figure} %aspect ratio: 1.3  word equivalent: 135

The degree of  {spin squeezing} is optimized by varying $\tau_\text{A}$ and $\tau_\text{B}$ as shown in Fig. \ref{fig: squeezdem} (b). For the squeezing preparation time  $\tau_\text{A}$, there is a clear optimum due to two opposing effects. For a too short $\tau_\text{A}$, the measurement strength limits the obtained information about the  {atomic spin} and thus the degree of squeezing. Extending  {$\tau_\text{A}$} too long leads to additional decoherence and depumping  {effects}.
The increase of $\tau_\text{B}$ beyond an optimal value degrades the level of squeezing due to the information loss by decoherence effects.  For the optimal values $\tau_\text{A} =  \SI{220}{\micro\second}$ and $\tau_\text{B} = \SI{40}{\micro\second}$, %corresponding to the balance between the measurement strength and induced decoherence, 
we observe $10\,\log\left(\xi^2\right) = \left(-4.6 \pm 0.6\right)$ dB of spin squeezing. Plotting the level of squeezing for $\tau_\text{B} = \SI{40}{\micro\second}$ versus the duration of $\tau_\text{A}$ for different duty-cycles $D$ of the stroboscopic pulses, we observe squeezing degradation due to worse back-action evasion (Fig. \ref{fig: squeezdem} (c)). 

The degree of  {spin squeezing} together with the atomic polarization allows us to estimate the degree of entanglement present in the macroscopic spin ensemble \cite{Sorensen2001}. With squeezing of $ \left(-4.6 \pm 0.6\right)$ dB and an initial atomic polarization of $> 0.97$, we find that the spin ensemble contains groups of up to ten entangled atoms. 
The non-negligible level of SN weakens the effect of the  {spin squeezing} on the overall observed noise.
For $\tau_\text{B}=\SI{40}{\micro\second}$ in Fig. \ref{fig: squeezdem} (d), SN and PN are of similar magnitude. Increasing $\tau_\text{B}$ to \SI{100}{\micro\second} improves the overall noise reduction in the conditional variance, enabling a more efficient measurement of a conductive sample at the expense of spin squeezing.

Next, we exploit  {spin squeezing} to demonstrate  {QMIT} with sensitivity improved beyond the projection noise limit.
For this, an RF pulse % is described in detail in supplements: fitting an integer number of oscillations at the Larmor frequency, corresponding to approximately 
of \SI{47}{\micro\second} duration is introduced in-between the two stroboscopic measurements (Fig. \ref{fig: signalVSphase} (a)).
 {For optimal measurement efficiency,
the stroboscopic pulses are required to} overlap with the cosine quadrature of the LIA reference signal. %we explain that extensively in the supplements. precise alignment of phases between the stroboscopic pulse train\blue{s}, the LIA reference signal and the RF pulse. For optimal readout, the stroboscopic pulse phases during both $\tau_A$ and $\tau_B$ are optimized to match the LIA cosine quadrature. %This can be done by slightly adjusting the delay of the probing sequence and the gap duration between two probing processes. UNNESSECARY as described in supplements and vey specific
%The phase of the atomic precession signal is determined by the RF pulse phase. To
Further, the eddy current detection is optimal when the RF pulse phase matches the LIA reference signal (Fig. \ref{fig: signalVSphase} (b)).

\begin{figure}[!]
\includegraphics[width=8cm]{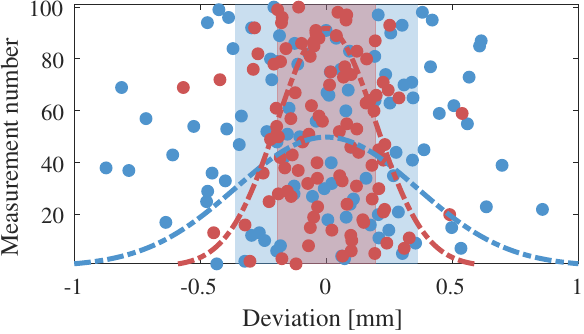}
\caption{Entanglement-enhanced 1D MIT. 
The statistical distribution of the sample center found from 1D MIT scanning 50 mm in 1 mm steps.  
One hundred red (blue) points corresponding to the results for the position of the center of the sample obtained with (without) spin entanglement. Each point is averaged over 40 measurement repetitions. 
The red (blue) shaded areas cover the red (blue) data points within one standard deviation. 
}
\label{fig: signalVSposition}
\end{figure} %aspect ratio: 1.75, word equivalent: 110

%\tauaoptimum{} and \taubmagnetometer{} (Fig. \ref{fig: signalVSphase}a) which prepare and detect the quantum state, 
The RF field %described previous in text: from the coils in the anti-Helmhotlz configuration (Fig. \ref{fig: exp. setup} (a), (b)) 
induces eddy currents in the sample, a small titanium piece of dimension $\SI{1}{\milli\meter}\times\SI{10}{\milli\meter}\times\SI{10}{\milli\meter}$. 
Fig. \ref{fig: signalVSphase} (b) visualizes the MIT signal from the sample as a function of the phase of the RF field. The signal is the difference between the sample and background measurement. 
The eddy current signal is maximal with an out-of-phase RF field. The respective uncertainties for conditional (red) and unconditional (blue) measurements  are shown in Fig. \ref{fig: signalVSphase} (b). For comparison, the quantum noise for the back-action evaded measurement without squeezing is shown in green. It is determined from the experimental results as $\sqrt{\text{SN}_\text{B}+\text{PN}_\text{B}}$, where $\text{PN}_\text{B}$ is the projection noise variance during $\tau_\text{B}$ and $\text{SN}_\text{B}$ {is} the shot noise {variance} as defined in Eq. (\ref{eq:squeezing}).
The average level of spin squeezing for the data shown in Fig. \ref{fig: signalVSphase} (b) is $\left(-1.8 \pm 0.1\right)\,$dB. %shown in Fig. \ref{fig: signalVSphase} b. 
The observed reduction in the level of squeezing originates from multiple factors. First, introducing a gap leads to a degradation of squeezing due to decoherence effects \cite{SM}. For a gap of  \SI{50}{\micro\second},  squeezing is reduced to $\left(-3.0 \pm 0.4\right)$\,dB when the RF coils are disconnected from electronic devices. 
The second effect reducing the available  squeezing is the connection of the RF coils to the function generator. While connecting the coils compromises the squeezing, we did not observe a significant change between sending the RF pulse or not. We therefore believe the degradation of squeezing originates from minuscule currents flowing through the coils even when no RF pulse is sent from the function generator. 
%In the following, we present those measures for the data shown in Fig. \ref{fig: signalVSphase} b obtained with $\left(1.8 \pm 0.1\right)\,$dB of spin squeezing.
Despite this, we still observe $\left(41 \pm 1\right)\,\%$  reduction between conditional to unconditional uncertainty%  amounts to $\left(41 \pm 1\right)\,\%$
. Considering the maximal signal at 90$^\circ$ RF phase, we observe 42.5\,\% noise reduction,  {improving the SNR of a single-shot measurement} from 0.72 to 1.2.
The entanglement-assisted sensing allows us to achieve a conditional uncertainty 11\,\% below the expected quantum noise for the back-action free measurement%for atoms in the CSS
, given as $\sqrt{\text{SN}_\text{B}+ \text{PN}_\text{B}}$.
This result matches well with the sensitivity improvement expected 
from the level of squeezing, estimated through $\sqrt{\text{SN}_\text{B}+ \xi^2\,\text{PN}_\text{B}}/\sqrt{\text{SN}_\text{B}+ \text{PN}_\text{B}}\approx 0.89$
\cite{SM}. 
Comparing the SQL of continuous measurements, given by  $(1+2/\sqrt{3})^{1/2} \approx 1.47$ in units of $\sqrt{\text{PN}_\text{B}}$ to our conditional noise of $1.11\,\sqrt{\text{PN}_\text{B}}$, the observed noise reduction can be estimated as $1.11 / 1.47 =0.76$, corresponding to %\purple{\sout{noise of}}
$-2.4$ dB noise reduction.% \purple{\sout{below/compared to the SQL}}.

Finally, we demonstrate the spatial sensitivity of our sensor with a one-dimensional
 (1D) QMIT of the sample.  {As shown in Fig. \ref{fig: exp. setup} (b)}, the sample is moved past the cell in 1 mm steps along the $x$-axis. 
%The sample is moved past the cell in 1 mm steps along the $x$-axis and
 {For each position,} 4\,000 consecutive measurements are performed. 
The  {conditional variance} is determined using $\alpha_\text{opm}=0.91$ from the no-sample measurement. 
 {We average} the sweep of the sample 40 times, corresponding to 100 independent  {MIT measurements}.
In Fig. \ref{fig: signalVSposition}, the distribution %the distribution of the location 
of the sample center for the 100 MIT measurements is shown with conditional and unconditional results marked in red and blue, respectively. 
The distribution is significantly narrower using spin squeezing, visualized using Gaussian distributions and shaded areas, reflecting the standard deviation of 0.20 mm for the conditional measurement and 0.36 mm for the unconditional measurement. The quantum-enhanced MIT provides a nearly two-fold improvement in precision. %The entanglement-enhanced measurement provides thus a nearly two-fold improvement of the precision of the 1D MIT. 

The duration of a single 1D tomography sequence can be estimated from the number of repetitions combined with the number of positions measured. Using a conservative estimate of 13 ms per repetition including optical pumping and the measurement, 40 repetitions per position would take 520 ms. The total scan sufficient to measure the sample position with an uncertainty of 0.20 mm  would take 26 s. 

We have proposed a novel quantum sensing protocol for magnetic induction tomography.  We successfully demonstrated entanglement-enhanced eddy current detection and 1D QMIT through back-action evasion and spin squeezing. The demonstrated improvement of sensitivity beyond the SQL offers a promising path towards non-invasive measurements on weakly conducting samples, such as biological tissue, exploiting the noise reduction for higher sensitivity and less measurement time.

%\end{document}
\section*{Acknowledgments}
The authors would like to thank Kasper Jensen for valuable input and discussion on both, experimental implementation and theoretical understanding.
The authors acknowledge many fruitful discussions with Jörg Helge Müller and Jean-Baptiste Beguin. Mikhail Balabas fabricated the alkene coated vapour cell used for this experiement.

We acknowledge funding by the Novo Nordisk Foundation grant NNF20OC0064182 within the "Exploratory Interdisciplinary Synergy Programme
2020", the EU grant MacQsimal, the European Research Council (ERC) under the EU Horizon 2020 programme (grant no. 787520), and the VILLUM FONDEN under a Villum Investigator Grant (grant no. 25880). W. Z. and H. W. acknowledge the support of the National Nature Science Foundation of China (grant no. 12075206).

% \bibliographystyle{naturemag}
 %\bibliography{bib2}

%%Updated PRL style bibliography do not outcomment the one below as it is new :) 
%

\end{document}

\begin{titlepage}

\centering

\vspace*{50 mm}

{\LARGE Supplementary Information \--- Entanglement-enhanced magnetic induction tomography}
\vspace{10 mm}

Wenqiang Zheng et. al.

\end{titlepage}

\clearpage

\onecolumngrid

\section{Experimental setup}

\begin{figure}[htpb]
\includegraphics[width=14cm]{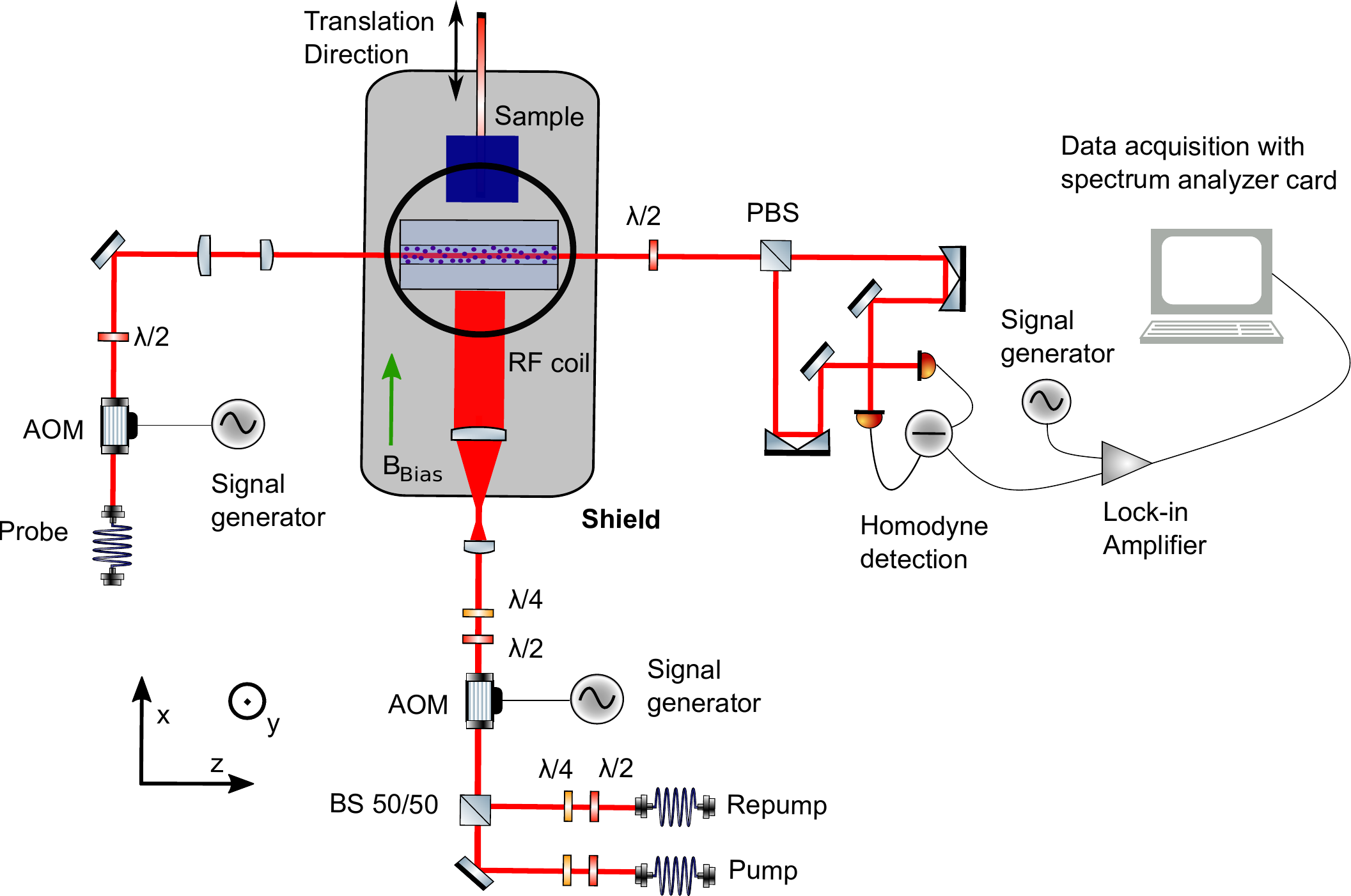}
\caption{Extended experimental setup. Experimental schematic including configuration of bias and RF magnetic field, vapour cell, sample and optical pumping and probe laser configuration. $\lambda/2$ and $\lambda/4$ are indicating half- and quarter-waveplates, (P)BS is indicating a (polarizing) beamsplitter, while AOM refers to an acousto-optical modulator. %\red{Snall we show stroboscopic pulses with duration, etc?}\blue{I tried incorporating it, but it does not look nice. Did a fast separate figure.}
}
\label{fig: experimentalsetupsupplements}
\end{figure}

\begin{figure}[H]
\centering
\includegraphics[width=12cm]{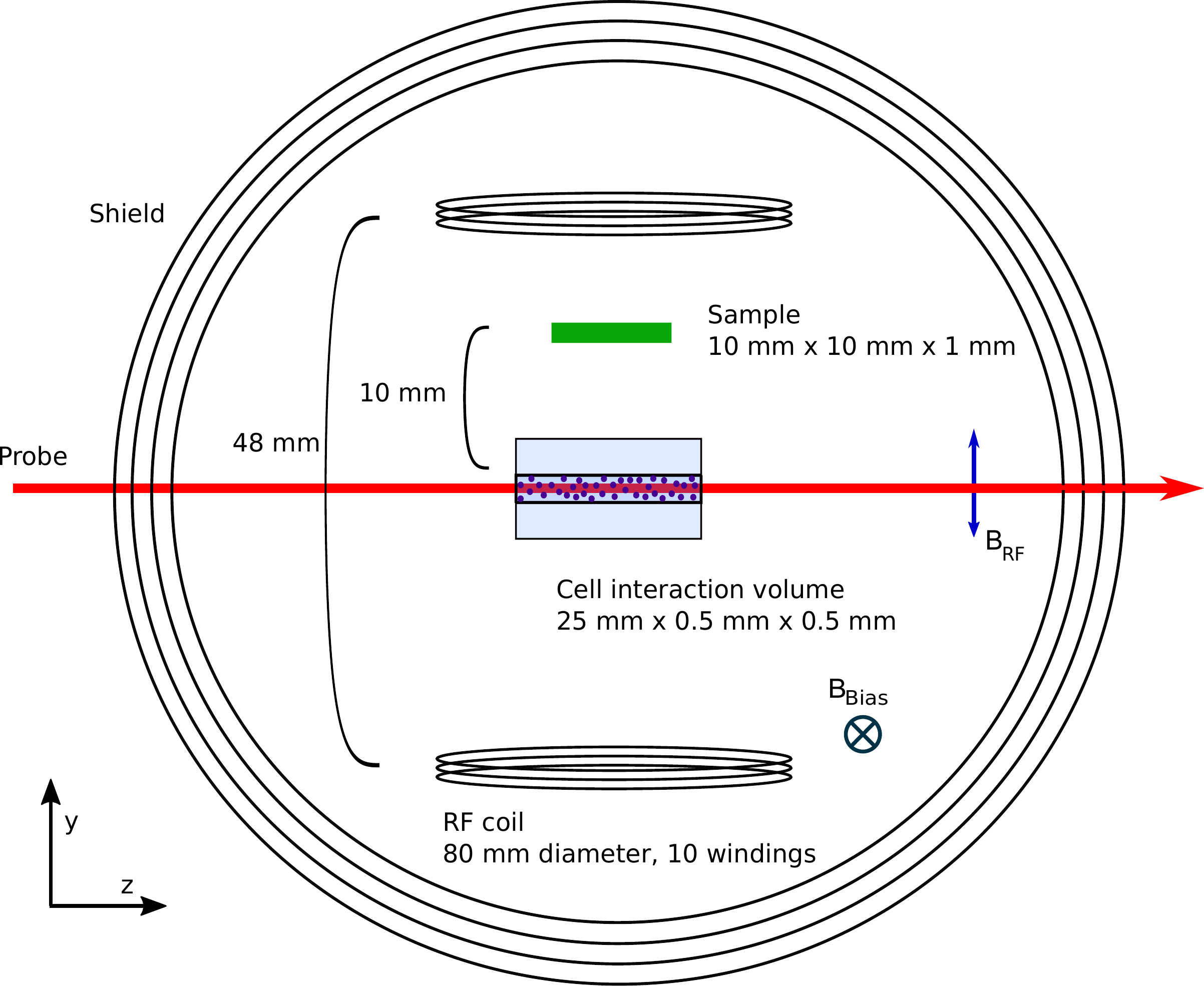}
\caption{Side view of coil- and cell configuration. Illustration of experimental geometry inside the magnetic shield including dimensions for the vapour cell, sample and RF coils used for the entanglement-enhanced MIT.
\label{fig: experimentalsetuptransversesupplements}}
\end{figure}

In the experiment, an encapsulated, anti-relaxation coated cesium vapour cell is used as the container of the atomic ensemble forming the atomic magnetometer (AM). The interaction volume is a micro channel of size  $\SI{500}{\micro\meter}\times\SI{500}{\micro\meter}\times\SI{25}{\milli\meter}$. The cell is placed inside a magnetic shield (Supp. Fig. \ref{fig: experimentalsetupsupplements} and \ref{fig: experimentalsetuptransversesupplements}) consisting of five layers, the outermost layer is made of iron, the three next inner layers consist of mu-metal and the innermost layer is made of aluminium.
Along the shield axis and transverse to the cell channel, a bias magnetic field $B_\text{bias}$ is applied, introducing a Zeeman level splitting corresponding to a Larmor frequency of $\Omega_\text{L} \approx 725$ kHz. 
Parallel to the bias magnetic field, optical pumping is used to prepare the atomic ensemble in the coherent spin state (CSS). We typically achieve an atomic polarization of > 97\,\% (Supp. \ref{Sec: statecalib}). 
Parallel to the cell channel and transverse to the macroscopic spin orientation, the stroboscopic probe beam, \SI{1.95}{\giga\hertz} detuned from the $F=4\rightarrow F'=$ 4,5 crossover transition of the D2-line, is sent. The pulses for the experimental sequence are generated by two acousto-optical modulators (AOMs), one for the stroboscopic probe pulses and one for the overlapped pump and repump laser as indicated in Supplementary Fig. \ref{fig: experimentalsetupsupplements}.
For the optical pumping, low-pass filtering of the drive signal sent to the AOM allows for a smooth turn-off of the optical pumping and therefore avoids unwanted atomic excitations due to the high-frequency components in a square pulse.

Transverse to both, probing and bias magnetic field direction, an optional RF field can be applied. For the atomic state characterization, the RF coils are connected in parallel. For MIT measurement, the differential technique (\cite{Jensen2019a}) providing zero mean field at the cell in the absence of any conducting sample is used. 
Eddy currents induced in the conducting sample inserted between one of the coils and the cell  (Supp. Fig. \ref{fig: experimentalsetuptransversesupplements}) change the mean field affecting the atomic spin state.
The atomic spin is mapped onto the probe light polarization which is recorded using homodyne detection. Since the measurement is phase sensitive, leading to artefacts in the recorded signal due to minor path differences, but also different response times of the two photo detectors, "cat-eyes" on  a stage are used to correct for any path length difference in the homodyne detection.
The signal is demodulated with a lock-in amplifier (LIA), the demodulated time-signal is acquired for thousands of consecutive repetitions. 
To maximize the detection efficiency, we optimize the demodulation phase of the LIA to match the stroboscopic phase.

\begin{figure}[H]
\centering
\includegraphics[width=12cm]{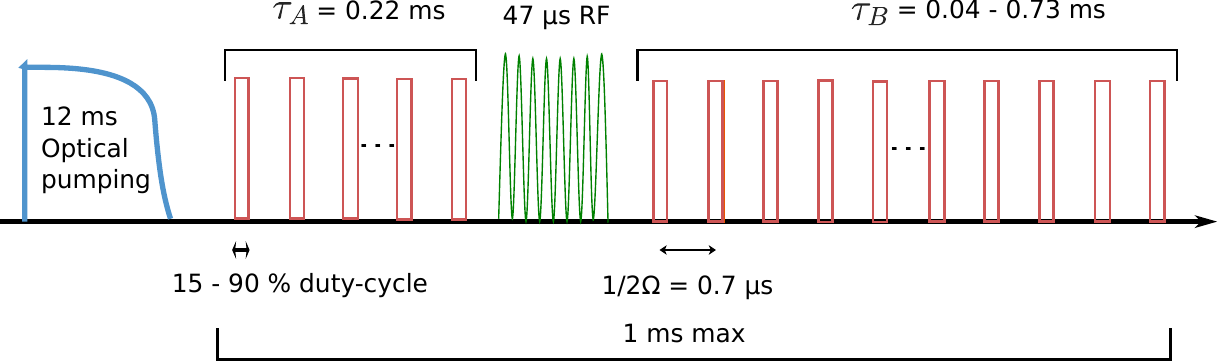}
\caption{Experimental time sequence for qMIT. The gap and RF pulse are optional, duration of \taub{} is truncated in the postprocessing analysis. The duty-cycle $D$ of stroboscopic pulse trains can be varied using the signal generator shown in Supp. Fig. \ref{fig: experimentalsetupsupplements}.
\label{fig: pulsessupplements}}
\end{figure}

All pulse sequences used have a rather long optical pumping pulse, exceeding the longitudinal spin coherence time $T_1$ several times, preparing a CSS as the initial state. The optical pumping is followed by a sequence of stroboscopic probe pulses, modulated at twice the Larmor frequency $\Omega_\text L$. The duty-cycle $D$ of the stroboscopic probing can be varied. 
The duration and strength of the consecutive probe pulses used for squeezing preparation \taua{} and squeezing verification \taub{} can be varied for squeezing optimization.  Introducing a gap of \taugap{} between the two processes allows to incorporate the RF pulse which generates eddy currents in the sample (Supp. Fig. \ref{fig: pulsessupplements}). The impact of the gap on the squeezing is described in \ref{sec: squeezdecay}.
%To incorporate the RF pulse for the MIT-type of measurement, a short gap of \taugap{} \blue{has to be} introduced between the two processes. 
%\blue{Applying an RF pulse generates eddy currents in the sample. %and only afterwards the second stroboscopic pulse sequence is applied to verify and retrieve the squeezing information. 

To gauge the strength of the RF field, the RF coils have to be calibrated. This is done using a pick-up coil of 9 mm diameter with 30 windings. First, the inductance of the pick-up coil is determined using a simple RL circuit and varying the frequency. The frequency dependent response is fitted to get the estimate for the inductance, here $L = \SI{30.5}{\micro H}$. Using this, the magnetic field versus applied voltage to the RF coils can be calibrated.

The performance of the entanglement-enhanced RF magnetometer is verified by inserting a small piece of titanium ($\SI{1}{\milli\meter}\times\SI{10}{\milli\meter}\times\SI{10}{\milli\meter}$) between one of the RF coils and the vapour cell (Supp. Fig. \ref{fig: experimentalsetuptransversesupplements}). Recording the change in the signal due to the induced eddy currents in the sample allows to detect the sample location and perform a one-dimensional (1D) MIT exploiting entanglement-enhanced sensitivity to eddy currents. %and the change in response due to the induced eddy currents is recorded.

%\blue{Outcommented MIT signal}
\begin{comment}

Moving the sample past the cell in steps of 1 mm, the eddy current signal can be recorded to perform a 1D MIT. In figure

\begin{figure}[H]
\centering
\includegraphics[width=10cm]{gausfit.pdf}
\caption{\textbf{Quantum-enhanced 1D MIT} Recorded 1D MIT signal for the sample moved past the cell in one mm steps. Each point reflects the mean of the recorded signal, while the error bars reflect the single-shot errors, where blue and red points reflect the unconditional and conditional uncertainty, respectively.
\label{fig: mitfit}}
\end{figure}

\end{comment}

\section{Atomic polarization and spin state calibration}
\label{Sec: statecalib}
A fundamental requirement for the successful quantum-noise-limited operation of the atomic ensemble as an AM is a high degree of atomic polarization, a prerequisite to the CSS. Optical pumping using a pump ($ F=4\rightarrow F' = 4$ on D1) and a repump ($F = 3\rightarrow F'=$ 2,3 crossover transition on D2) laser is employed for a duration exceeding the spin relaxation time many times. To quantify the atomic polarization of the atomic ensemble, the technique of magneto-optical resonance spectroscopy (MORS), also referred to as RF spectroscopy, as presented in \cite{Julsgaard2004}, is used. Given that the experiment is operated using laser pulses, the MORS is also performed in its pulsed version. 
Due to the RF pulse's short duration, it is spectrally broad and hence is driving all oscillations between neighboring Zeeman levels despite their different Larmor frequencies due to the quadratic Zeeman splitting.

\begin{figure}[H]
\centering
\includegraphics[width=15cm]{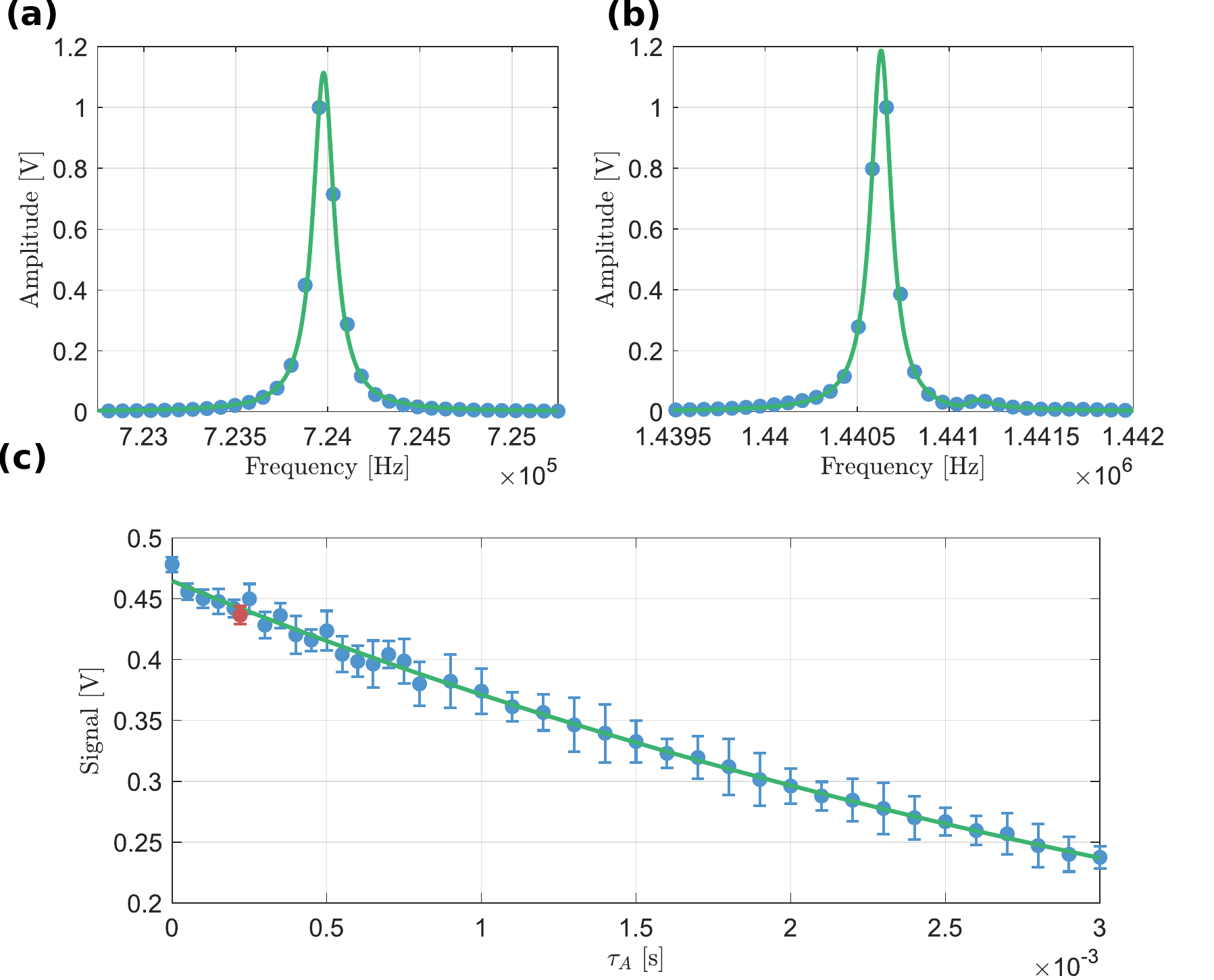}
\caption{Atomic state calibration and CSS decay. (a) Unresolved pulsed MORS spectrum at Larmor frequency $\Omega_\text L \approx \SI{724}{\kilo\hertz}$ used for the quantum-enhanced AM.
(b) Resolved pulsed MORS spectrum at $\Omega_\text L \approx \SI{1.4}{\mega\hertz}$ with the fit for estimating atomic polarization. A small maximum around 1.4412 MHz is a signature of imperfect spin polarization.
(c) Signal amplitude during \taubmagnetometer{} versus duration \taua{}. MORS spectra are measured after squeezing generation process. The error bars reflect the standard deviation on the mean obtained from 4000 measurements for each point. The red data point is the signal amplitude for the optimal squeezing for \tauaoptimum{}. The fit is an exponential decay.
}
\label{fig: pmors}
\end{figure}

%\purple{\sout{For the operation as an AM, }}\purple{
The bias magnetic field $B_\text{bias}$ is chosen such that the Larmor frequency is $\Omega_\text{L} \approx \SI{725}{\kilo\hertz}$. At the operating temperature of $55\,^\circ$C the broadening arising from atom-wall collisions and magnetic field inhomogeneity does not allow for a resolved pulsed MORS spectrum (Supp. Fig. \ref{fig: pmors} a) and hence it is not possible to estimate the atomic polarization reliably. In order to get a proper estimate of the atomic polarization, the Larmor frequency is increased to $\Omega_\text{L} \approx \SI{1.4}{\mega\hertz}$ to increase the quadratic splitting beyond the broadening and to resolve the spectrum. The measured atomic polarisation in this case is 97.5\,\% (Supp. Fig. \ref{fig: pmors} b). From this we infer the atomic polarization after the optical pumping to be $> 97$\,\%  for the usual Larmor frequency. This high level of atomic polarization is obtained through tedious optimization of the circular polarization, beam size and power, as well as shaping the turn-off of the optical pumping beams to be smooth.

From the pulsed MORS shown in Supp. Fig. \ref{fig: pmors}  b, only the polarization directly after the optical pumping is obtained. 
However, since the experimental pulses extend over multiple hundreds of microseconds, the atomic polarization decays during the measurement sequence. To quantify the influence of the decay of the CSS during the squeezing preparation, the duration of the first stroboscopic pulse has been varied before sending a RF pulse and  the signal has been consecutively read out for a fixed duration of \taubmagnetometer{}. %\sout{Since the experiment cannot be operated at 1.4 MHz, this poses a problem as we cannot simply map out the atomic state after sending the stroboscopic pulses during \taua{}. Therefore, an alternative way of quantifying the influence of the decay from the CSS during the squeezing preparation has to be used. For this, the duration of the first stroboscopic pulse has been varied before sending a RF pulse \taugap{} and consecutively reading out the signal for a fixed duration of \taubmagnetometer{}.} 
Depending on the decay of the atoms from $m_F=4$ of the hyperfine level $F=4$ into other Zeeman levels or into the hyperfine level $F=3$, the strength of the signal should decrease. The decay rate should be directly related to the the atomic population distribution and hence be given by 
\begin{align}
    A(t) \qquad = \qquad B\exp{\left(-t/T_1\right)}.
\end{align}
Fitting this exponential decay to the data as presented in Supplementary Fig. \ref{fig: pmors} c) leads to a spin decay time $T_1$ of
\begin{align}
    T_1 \qquad = \qquad (4.5 \pm 0.1)~\SI{}{m\second}.
    \label{eq: fitresultt1}
\end{align}
This is significantly longer than the duration of \taua{}, therefore the impact arising from polarization decay should be small.
%. After \tauaoptimum{}, \blue{before continuing we should be sure that these are claims we want to make.}

\section{Spin noise calibration}
\label{Sec: statecalib}

The measured spin noise of the unpolarized, thermal atomic ensemble, namely thermal spin state (TSS) is considered a robust reference for calibrating the spin noise of the CSS, because the TSS is insensitive to the classical noise and BAN. For the TSS, all 16 sublevels of the ground state have the same population, so there are 9/16 of all atoms in the F = 4 state. Remaining atoms staying in the F = 3 state are not observed by the probe light due to a large probe detuning. The spin noise variance of a single atom in the hyperfine state F = 4 is 
$\text{Var}( \hat{j}_z )_\text{TSS} =\left< \hat{j}_{z}^{2} \right> ={\left< \hat{j}^2 \right>}/{3}={F( F+1 )}/{3}={20}/{3}$, so the observed noise for the TSS is $\text{Var}( \hat{J}_z )_\text{TSS} = \frac{20}{3}\times \frac{9}{16}N_\text{A}=\frac{15}{4}N_\text{A}$, whereas for the CSS, the spin noise is $\text{Var}( \hat{J}_z )_\text{CSS}=2N_\text{A}$. Therefore, we use $\text{Var}( \hat{J}_z )_\text{CSS}=\frac{8}{15}\text{Var}( \hat{J}_z )_\text{TSS}$ to calibrate the spin projection noise of the CSS using the observed spin noise of the TSS.

In the experiment, we calibrate the projection noise variance $\text{PN}_\text{B}$ with reference to TSS. For this, the underlying assumption is that the atomic ensemble is in a CSS state. 
However, following the characterization of the atomic state presented in Supp. Fig. \ref{fig: pmors} a and b, it becomes clear that this assumption does not perfectly hold, since we achieve 97.5\,\% atomic polarization after optical pumping. 
Since the CSS has the smallest projection noise, the calibration used is underestimating the true projection noise. 
From the 97.5\,\% of atomic polarization, the calibration of the projection noise correction for the initial spin state amounts to 20\,\%.  %\textbf{This is higher than the 17 \% from Qiang, my assumption is that he used a different atomic polarisation?} 
%Comment: Now it depends what we compare exactly. I chose to pick the data for the 90 degree RF phase in the following, however the changes due to some 2nd or 3rd digits is minor.
For the sample measurement, the improvement due to squeezing can be calculated as the ratio of the eddy current sensitivities with and without the squeezing. For the uncorrected projection noise (see the main text), the improvement is
\begin{align}
    \sqrt{\text{SN}_\text{B}+ \xi^2\,\text{PN}_\text{B}}/\sqrt{\text{SN}_\text{B}+ \text{PN}_\text{B}}=\frac{ \sqrt{1 + 1.75/10^{0.18}}}{\sqrt{1 + 1.75}} \quad = \quad 0.89 
\end{align}
which corresponds to a reduction of 11 \% in the uncertainty compared to the projection noise without correcting for the imperfect atomic polarisation.
Considering the imperfect spin polarization, which leads to correction $\text{PN}_\text{B}\rightarrow1.2 \text{PN}_\text{B}$, the ratio changes to 0.88. This is only a small change to the uncorrected value.

\section{Verification of back-action evasion}
In the experiment stroboscopic pulses are employed to reduce the amount of back-action introduced to the system. While complete back-action evasion is expected only for $\delta$-peak like interaction, meaning an infinitely short probe pulse, experimentally there are limits to the applied duty-cycle of the probe pulse.
In particular, the AOM used for generating the stroboscopic probing pulses puts a limit on the duty-cycle to 15\,\%.

\begin{figure}[H]
\centering
\includegraphics[width=9cm]{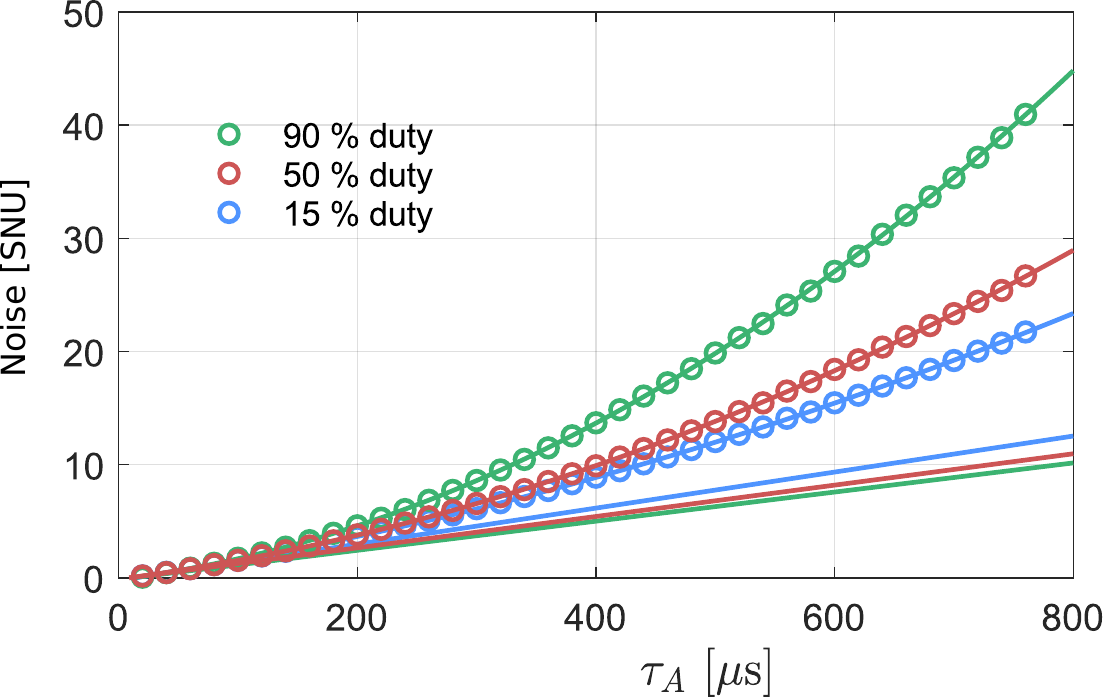}
\caption{Verification of the back-action evasion. Reduction of back-action noise for lower duty-cycles: 90\,\% (green), 50\,\% (red) and 15\,\% (blue). The linear contribution represent the projection noise, while the data points contain projection and back-action noise.}
\label{fig: backaction evasion}
\end{figure}

The cosine quadrature of the LIA is determined by 
\begin{equation}
\label{DutyNoise}
\text{Var}\left( \hat{S}_{y,\tau}^{\text{out}} \right) \approx \frac{\Phi \tau}{{4}}\left[ 1+\sinc\left( \pi D \right) \right] \left[ 1+{\frac{\tilde{\kappa}^2}{2}}+{\frac{\tilde{\kappa}^4}{12}}\frac{1-\sinc\left( \pi D \right)}{1+\sinc\left( \pi D \right)} \right],
\end{equation}
(see the main text and \cite{Vasilakis2015b}) where $D$ is the duty-cycle. In order to compare the quantum noise and verify the reduction in the back-action noise for different duty-cycles, the obtained signal has to be corrected for the different overlap with the LIA demodulation, corresponding to dividing the recorded signal by $\eta=\left[ 1+\sinc\left( \pi D \right) \right]$. The respective signals for duty-cycles of 90, 50 and 15 \% are shown in Supp. Fig. \ref{fig: backaction evasion}.
As the same average power per duty-cycle is used, the linear contribution arising from the projection noise in equation (\ref{DutyNoise}) should be the same for all duty-cycles after the correction for $\eta$. However, given the different back-action contributions, which is quadratic, due to the different duty-cycles, these should not coincide. In Supp. Fig. \ref{fig: backaction evasion} it is clearly visible that the back-action contribution is lower for smaller duty-cycles, while the linear part obtained from fitting to a polynomial of second order as suggested by equation (\ref{DutyNoise}).

As seen in Supp. Fig. \ref{fig: backaction evasion}, the projection noise level estimated from the fit for a duty-cycle of 15\,\% (blue line),  does not coincide exactly with the lines for the other two duty-cycles. This seems to be related to the fact that the AOM has a finite rise time and the approximation of a square pulse starts to break down. 

\section{Squeezing decay}
\label{sec: squeezdecay}
For the operation as an atomic magnetometer, the pulses for squeezing generation and verification cannot be sent consecutively, but rather a time gap that allows for sending an RF pulse has to be introduced. 

\begin{figure}[]
\centering
\includegraphics[width=11cm]{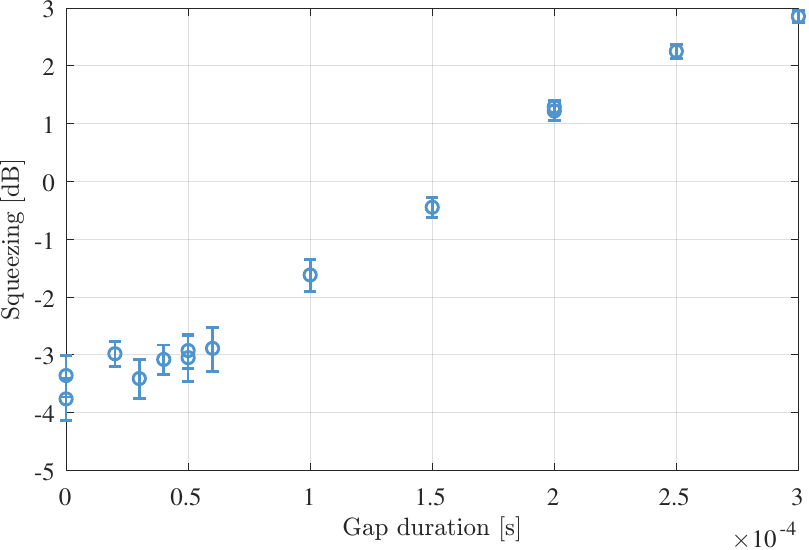}
\caption{Squeezing degradation versus gap duration. Level of achievable squeezing in dB for fixed duration of squeezing generation $\tau_A=\SI{220}{\micro\second}$ and verification $\tau_B=\SI{100}{\micro\second}$ versus the gap duration between the first and second stroboscopic pulse. The error bars are obtained from 9 independent data sets containing 4000 experimental repetitions each. Gaps with multiple points were taken on different days to show comparability and repeatability of the measurement performance. }
\label{fig: squeezvsgap}
\end{figure}
To test the effect of a gap on the level of squeezing, a stroboscopic pulse of duration  \tauaoptimum{} for the squeezing preparation with a 15\,\% duty-cycle is sent. The optimum duration has been determined from data without a gap as presented in Fig. 2 b, c of the main text. 
Delaying the verification pulse \taub{} in a range between 0 - 300 \SI{}{\micro\second} allows to estimate the influence of spin decoherence in the dark on the level of spin squeezing. In Supp. Fig. \ref{fig: squeezvsgap}, the obtained level of squeezing for a fixed duration of $\tau_\text{B}=\SI{100}{\micro\second}$ is shown. In the experiment itself the duration of \taub{} can be chosen in the analysis by truncating the time bins considered for the analysis. Given the trade-off between SNR and degree of squeezing, we chose to fix $\tau_\text{B}=\SI{100}{\micro\second}$ for Supp. Fig. \ref{fig: squeezvsgap}. With the gap between the two stroboscopic pulses the squeezing degrades. In Supp. Fig. \ref{fig: squeezvsgap} the degradation of the squeezing is small for short gap duration less than \SI{60}{\micro\second}, but the decay of the squeezing is more severe for longer gap durations. Given the trade-off between the squeezing degradation and the possibility to drive excitations and eddy currents for the quantum-enhanced MIT type of measurement, the gap duration is chosen to be \SI{50}{\micro\second}.

It should be noted that in Supp. Fig. \ref{fig: squeezvsgap} there are two points for 0, 50 and \SI{200}{\micro\second} gaps, originating from data sets acquired on different days. This was to verify that the results are reproducible. The error bars in the plots originate from nine individual measurements of the thermal noise, atomic noise, the photon shot noise and the electronic noise, each consisting of 2\,000 measurement repetition. The obtained variances and their uncertainties are then propagated through using the standard error propagation formula as was done for all other measurements for the degree of squeezing.

\section{Influence of F=3 population on the projection noise} 
In the experiment, a probe laser at $\varDelta' =\SI{-1.95}{\giga\hertz}$ blue detuning from the $F=4\rightarrow F'=$ 4,5 crossover transition of the D2-line is used. While most atoms should be in the $F=4$ hyperfine manifold, residual atoms in the $F=3$ can contribute to the obtained signals as well. Therefore, the expected ratio of their contributions should be estimated.

The measured noise is expressed by equation (\ref{DutyNoise})
%\cite{Vasilakis2015b}
%\begin{align}
%\text{Var}\left[ \hat{S}_{y,\tau}^{\text{out}} \right] \approx \frac{\Phi %\tau}{8}\left[ 1+\sinc\left( \pi D \right) \right] \left[ &1+\hat{\kappa}^2+\frac{\hat{\kappa}^4}{3}\frac{1-\sinc\left( \pi D %\right)}{1+\sinc\left( \pi D \right)} \right],
%\end{align}
with $\hat{\kappa}^2=\frac{1}{4}\beta ^2J_x\Phi \tau \left[ 1+\sinc\left( \pi D \right) \right]$, $\beta =-\frac{\Gamma}{8A\Delta}\frac{\lambda ^2}{2\pi}{a}_1$. Here, the detuning $\Delta$ and Faraday interaction strength $a_1$ depend on whether atoms are in $F=4$ or $F=3$. 
For the Cesium D2 transition ($6^2S_{1/2}\rightarrow 6^2P_{3/2}$) from the $F=4$ hyperfine manifold of the ground state \cite{Julsgaard2003}, 
$$
a_1^{F=4}=\frac{1}{120}\left( -\frac{35}{1-\varDelta _{35}/\varDelta}-\frac{21}{1-\varDelta _{45}/\varDelta}+176 \right), 
$$
where $\varDelta _{35}= 452.24~\SI{}{\mega\hertz}$, $\varDelta _{45}= 251.00~\SI{}{\mega\hertz}$ \cite{Steck2009} and $\varDelta =\SI{-1.82}{\giga\hertz}$ with respect to transition $6^2S_{1/2}~F=4\rightarrow 6^2P_{3/2}~F'=5$. So for $F=4$, $a_1\approx 1.079$. The unperturbed ground state hyperfine transition frequency of the cesium-133 atom is \SI{9.193}{\giga \hertz}. Similarily, $a_1$ can be estimated for atoms in $F=3$ \cite{Julsgaard2003}
\begin{align}
    a_1^{F=3} = \frac{1}{56}\left( \frac{45}{1+\varDelta _{24}/\varDelta}-\frac{21}{1+\varDelta _{23}/\varDelta}-80 \right), 
\end{align}
with $\varDelta =\SI{+6.76}{\giga\hertz}$ with respect to transition $6^2S_{1/2}~F=3\rightarrow 6^2P_{3/2}~F'=2$. Then $a_1\approx -1.032$ for $F=3$. 

The projection noise contribution $
\text{Var}\left[ \hat{S}_{y,\tau}^{\text{out}} \right]  \propto \hat{\kappa}^2\propto \beta ^2\propto a_{1}^{2}/\varDelta ^2$. For atoms in $F=3$ this ratio is 15 times smaller that for atoms in $F=4$. Given that optical pumping leaves only a minor part of atoms in $F=3$, their contribution to the signal should be even smaller. Therefore neglecting contributions to the signal arising from atoms in $F=3$ is a valid assumption.

%\blue{Using expression of Julsgaard for F=4 and blue-detuing being negative, I obtain for -1.95 MHz (our Probe detuning),a1 = 1.07576 using the transitions from Steck, while the equation is 5.16. This also fits the plot that for Delta-infinity the a1 goes to 1 and before it is higher. The 0.882 do not fit this. As I am unsure about the location of the 4-5 cross-over, there might be a correction of 100 MHz necessary, but this correction is small. For the 3 manifold, the expresission is slightly different, however we should use the expression in the erratum, which leads to -1.0322. With both of these two values we obtain a ratio of 14.98. It should be noted that we do not include the factor of 2 pi, as the 9 GHz also do not include it and they would fall out in the expression anyway.}

%Wolframexpression for a1 (F=4): 1/120(-35/(1-(263.81+188.44)*10^6/(-1.95*10^9))-21/(1-(263.81)*10^6/(-1.95*10^9))+176)

%For F=3: 1/56(45/(1+(151.21+201.24)*10^6/((-0.26381-0.33964+9.1926-1.95)*10^9))-21/(1+(151.21)*10^6/((-0.26381-0.33964+9.1926-1.95)*10^9))-80)

\bibliography{bib2}